# DIVERSE R-PPG: CAMERA-BASED HEART RATE ESTIMATION FOR DIVERSE SUBJECT SKIN-TONES AND SCENES

Authors: Pradyumna Chari[,1], Krish Kabra[1], Doruk Karinca[1], Soumyarup Lahiri[1], Diplav Srivastava[1], Kimaya Kulkarni[1], Tianyuan Chen[1], Maxime Cannesson[2], Laleh Jalilian[2] and Achuta Kadambi[1,3,*]

**Affiliations:** [1] *Department of Electrical and Computer Engineering,* [2]*Department of Anesthesiology and Perioperative Medicine,* [3] *California NanoSystems Institute, University of California, Los Angeles, USA*

*Correspondence: achuta@ee.ucla.edu

**ABSTRACT**:

Heart rate (HR) is an essential clinical measure for the assessment of cardiorespiratory instability. Since communities of color are disproportionately affected by both COVID-19 and cardiovascular disease, there is a pressing need to deploy contactless HR sensing solutions for high-quality telemedicine evaluations. Existing computer vision methods that estimate HR from facial videos exhibit biased performance against dark skin tones. We present a novel physics-driven algorithm that boosts performance on darker skin tones in our reported data. We assess the performance of our method through the creation of the first telemedicine-focused remote vital signs dataset, the VITAL dataset. 432 videos (~864 minutes) of 54 subjects with diverse skin tones are recorded under realistic scene conditions with corresponding vital sign data. Our method reduces errors due to lighting changes, shadows, and specular highlights and imparts unbiased performance gains across skin tones, setting the stage for making medically inclusive non-contact HR sensing technologies a viable reality for patients of all skin tones.

**INTRODUCTION**

Heart rate (HR) is an important clinical measure in the evaluation of cardiorespiratory and hemodynamic stability. Conventional HR assessment is performed in-person at a clinic or hospital using specialized monitoring equipment. However, the COVID-19 pandemic has accelerated the adoption of healthcare delivery to a remote model that uses telemedicine and mobile technologies for patient evaluations [1–3]. The assessment of HR in patients with suspected COVID-19 is particularly important, as COVID-19 has been associated with pre-existing cardiovascular disease [4]. Given the clinical relevance of HR in triage decisions, diagnosis, prognosis, and as a criterion for transfer to higher-level medical care, there is a pressing need to develop HR sensing solutions that can facilitate telemedicine-based care and remote patient monitoring in order to protect patients and healthcare workers from infectious exposure in a pandemic setting.

Recent methods have proposed using camera-based hardware in combination with computer vision algorithms and artificial intelligence (AI) tools to estimate HR in a completely contactless manner. Previous works have primarily focused on techniques that remotely extract a blood volume pulse (BVP) signal and corresponding HR estimate [5–38]. Remote photoplethysmography (r-PPG) is one of the most promising techniques used to extract a BVP, primarily from the face. r-PPG operates by looking for subtle color variations visible on the surface human skin, caused by sub-dermal light absorption fluctuations from changes in blood volume and content. Several r-PPG algorithms have been proposed to extract the BVP signal from videos, including blind source separation (BSS) [5,24,25], model-based [7,13,21,26], unsupervised data-driven [10,22], and supervised deep learning [17–19,23,27,35] methods. Unfortunately, the performance of existing algorithms fluctuates with changes in illumination condition [29], subject motion [12,26,30], and skin tone [39]. These key issues suggest that current r-PPG algorithms may be inherently *biased*: a performance gap exists for certain types of skin tones, subject motions (e.g. speaking), or illumination conditions.

Addressing these biases is essential for successful deployment of r-PPG technology in telemedicine applications, yet it remains a challenge. For example, dark skin, which contains higher amounts of melanin, fundamentally reduces the signal to noise ratio of all existing r-PPG algorithms. The important work of Nowara et al. [39] highlights this reduction, thereby conclusively determining that current r-PPG algorithms have markedly worse performance on darker skin tones. The work also highlights the issue of biased skin tone and gender representation in computer vision datasets, which is especially true for the comparatively small datasets used in r-PPG analyses. This dataset bias further prevents underlying algorithmic biases, such as skin tone bias, from being addressed. This is significant, since it has also been established that through the COVID-19 pandemic, the African American community in the United States have been the most affected [50,51]. These communities have also been found to have a higher than average prevalence of cardiovascular diseases [59]. Therefore, should non-contact HR sensing using video be implemented in a clinical setting, the development of r-PPG computer vision algorithms and datasets that improve the accuracy and reduce the bias of HR measurements for patients of all skin tones (especially the darker skin tones) is critically necessary for high-quality telemedicine care.

In this paper, we provide a novel approach at mitigating bias for skin tone. Kumar et al. (DistancePPG) [11] first attempted to reduce skin tone bias using a weighted average of signals from various facial regions-of-interest (ROI). However, to the best of the authors' knowledge, no work yet has continued development of r-PPG algorithms that tackle the important issue of performance bias on darker skin tones.  In contrast to prior approaches, the focus of this work is on understanding the unique physics that underlies inconsistency in r-PPG measurement. Using physics-rooted knowledge and camera noise analysis, we propose modifications to existing r-PPG denoising methods that use a similar weighted ROI philosophy as in DistancePPG (c.f. [8,9,32]). To assess the performance of the proposed method without dataset bias, we collect the first remote vital signs detection dataset focused on telemedicine applications that is demographically diverse. As a primary outcome measure of this paper, we qualitatively and quantitatively compare the performance of the proposed method against two popular categories of r-PPG algorithmic processing steps across varying skin tones and recording conditions. As a secondary outcome measure, we look at performance gain across skin tones and recording conditions in order to assess how the proposed method bridges the performance gap that is prevalent in current r-PPG algorithms.

**RESULTS**

**The VITAL Dataset**

In order to validate the performance of camera-based vital sign detectors, we construct the *Vital-sign Imaging for TeLemedicine AppLications* (VITAL) dataset. The focus of this dataset is to represent diversity in factors that are relevant to telemedicine setups, including: (i) smartphone deployment, (ii) camera view angle, (iii) recording condition diversity (lighting variation and talking), and (iv) patient demographic diversity. We address each of these aspects individually to highlight the extent of diversity in the dataset and how it was achieved:

*(i) Smartphone deployment:* The ubiquity of smartphones globally has led to the development of patient portals, many of which can be accessed via smartphone applications that can be downloaded by patients [40–42]. Such applications have been used for hosting telemedicine appointments. A deployable remote HR estimation solution with a focus on telemedicine must be able to work efficiently on smartphone cameras by considering factors including video compression [43] and algorithmic time and space complexity. Moreover, the solution must achieve success independent of camera type. In order to allow for such testing, the VITAL dataset uses different smartphone cameras for each view angle. The use of more than one smartphone imager inspires the development of algorithms that can scale to a variety of device-agnostic, telemedicine conditions.

*(ii) Camera view angle:* In a telemedicine setting, there can also be a diversity of camera angles that the algorithm must work on. In order to facilitate this estimation and verification, the VITAL dataset consists of two camera view angles for all the videos of each subject (as seen in Figure 1).

*(iii) Recording condition diversity:* Another essential factor involves testing algorithms across a range of recording conditions. The VITAL dataset consists of four recording conditions: (1) controlled lighting at 5600K ("cool" lighting) with the subject remaining stationary, (2) controlled lighting at 3200K ("warm" lighting) with the subject remaining stationary, (3) ambient room lighting- with distributed white LED lighting- with the subject remaining stationary, and (4) ambient room lighting with the subject speaking. As the background could not easily be varied, a green screen backdrop is kept to potentially enable digital modification of background scenery. The motivation for collecting these varied scene conditions in VITAL was to promote the development of telemedicine algorithms that can operate in the wild.

*(v) Patient demographic diversity:* The VITAL dataset consists of 54 subjects spread across skin tone, age, gender, race, and ethnic backgrounds. Subject characteristics (gender, age, height, weight, body mass index (BMI), race, and ethnicity) are summarized in Table 1 using mean (SD), median (IQR), or frequency (%), unless otherwise noted. For the purpose of this study, we split the subjects into three skin tone categories based on the Fitzpatrick (FP) skin type scale [44]: light, consisting of skin tones in the FP 1 and 2 scales, medium, consisting of skin tones in the FP 3 and 4 scales, and dark, consisting of skin tones in the FP 5 and 6 scales. This aggregation allows for more relevant trends, since any two consecutive FP scale categories are reasonably close; further, aggregation results in more objectivity of tests, given the qualitative nature of the Fitzpatrick scale.

**Benchmark Methods and Techniques**

To benchmark the performance of the proposed method, we compare the proposed method against previous remote HR estimation algorithms. We choose the CHROM [7] signal extraction method due to its versatility and open availability of code [45]. We compare with the two most common categories of algorithmic processing steps, which we refer to as *facial aggregation* (c.f. [5,7,10,13,25,26]) and *SNR weighting* (c.f. [8,9,11,32]). Both these techniques are described in detail in the Methods section. We believe that these two processing steps regimes encapsulate the major processing philosophies used in existing r-PPG methods.

To ensure a fair comparison with the benchmark methods, we implement identical testing conditions across techniques. Hence, for each method, the input video is passed through the same face detection algorithm (convolutional neural network based detector [46]), following which the eyes and mouth are cropped out using facial feature points [47]. Some methods also use skin segmentation algorithms to remove regions such as eyes and mouth [16,30,33], but we empirically found this to perform slightly worse on the VITAL dataset. We also use a consistent HR selection technique for each method, which also consists of a compression artifact suppression step. This is detailed in the Methods section.

**Results Summary**

Table 1 describes the distribution of subjects across various demographic metrics. Overall, remote HR estimation performance was compared across 54 subjects, across 4 scene conditions and 2 camera angles, resulting in a total of 432 videos with an average length of 2 minutes. HR estimation is carried out for windows of duration 10 seconds, with an overlap of 5 seconds. The overall HR for the subject is then estimated by averaging these window-estimated HR. Table 2 contains a performance summary across all statistical metrics employed- namely the Mean Absolute Error (MAE), Standard deviation of the error (SE) and the correlation coefficient (r) (details in the Methods section). In addition, Table 3 contains information about improvement in the Mean Absolute Error (MAE) metric for the SNR weighting and proposed methods, over the facial aggregation method.

The experiments highlight that, primarily, the proposed method: (i) shows an overall performance increase on the skin tone diverse VITAL dataset, (ii) shows debiased performance gain across skin tones, which is shown to not be the case with existing methods, (iii) is robust to recording conditions such as lighting, and (iv) is robust to camera placement with respect to the subject. Secondary observations include the nature of bias in existing methods, and the nature of performance differentials across scene conditions and camera angles. These aspects are discussed in some detail in the following subsections.

**Overall Performance**

Figure 2 shows the qualitative performance of the proposed method in comparison to the ground truth PPG and benchmark methods. The estimated pulse volume signal for the proposed method is found to visually contain peaks at the same frequency as the ground truth PPG signal. In some instances, the dicrotic notch is also present, although less prominent. Particularly noisy regions of the video are highlighted by the dashed red lines. In these time windows, the proposed method is found to visually recover peaks more distinctly with less high frequency artifacts in comparison to the benchmark r-PPG methods. Additionally, Figure 2b shows the beat-to-beat time evolution of the HR estimate, across the 10 second windows. Both the estimates from the ground truth signal and the output of the proposed method follow similar trends, consistently staying within 5 bpm of each other. However, because of the high frequency artifacts in existing methods, the estimated HR suffer from large errors in localized regions, worsening the overall HR estimate across the 2-minute video. Such qualitative improvements also translate quantitatively, where the proposed method shows a sub-6 beats per minute MAE for all skin tones (only method able to achieve this on the VITAL dataset), with an overall average MAE of 4.17 beats per minute.

**Skin Tone Performance**

For all three methods, performance degrades from light to dark skin. The facial aggregation approach observes a performance of 3.94, 4.14 and 6.20 beats per minute (bpm) for light, medium and dark skin tone subjects, resulting in an overall average performance of 4.49 bpm. The SNR weighting approach shows an improvement of 0.08 bpm in terms of MAE as compared to the facial aggregation benchmark, for light skin tones. However, the performance successively degrades as skin tone gets darker- medium skin tones are worse by 0.31 bpm, while dark skin tones are worse by 1.04 bpm, a significant drop. Hence, on a skin tone diverse dataset such as ours, this leads to a *decrease in overall performance*, with an average performance of 4.81 bpm. This is worse off by 0.32 bpm on average as compared to facial aggregation. Note the significant drop in the correlation coefficient (r) for dark skin tones, which goes from 0.44 for facial aggregation to 0.30 for SNR weighting.

In contrast, the proposed method shows significant improvement across all skin tones. That is, an improvement of 0.20 bpm, 0.31 bpm and 0.55 bpm for light, medium and dark skin tones respectively, in comparison to the facial aggregation benchmark. As a result of these significant improvements across all skin tones, overall performance on the dataset improves to a MAE of 4.17 bpm, an improvement in error performance of 0.32 bpm. As with previous methods, the performance of the proposed method is best for the light skin tone and reduces with darker skin tones; however, as compared to the benchmarks the proposed algorithm boosts the overall performance (across all skin tones), and the performance of darker skin by a larger amount. The proposed method does not obviate skin tone bias but rather is the first work that can be demonstrated to mitigate skin tone bias in the VITAL dataset.

Figure 3 highlights the high correlation between the proposed method's r-PPG HR estimates and ground truth PPG HR for light (r = 0.83) and medium skin tones (r = 0.85), and moderate correlation for dark skin tones (r = 0.52). The B&A plots in Figures 3d-f show a less than 2 bpm bias across all skin tones, and that all the proposed method's r-PPG HR estimates are mostly within 10 bpm of the ground truth. These correlation metrics are an improvement to the benchmark methods of facial aggregation and SNR weighting. Supplementary Fig. 1 and Supplementary Fig. 2 show the corresponding scatter and B&A plots for the facial aggregation and SNR weighting methods respectively.

**Recording Condition Performance**

Each of the three methods performs similarly across the three lighting conditions. The facial aggregation method shows an average MAE of 4.05 bpm across the lighting conditions, while the SNR weighting method shows an average performance of 4.46 bpm. This represents a decrease in performance of 0.41 bpm on average across the three lighting conditions. In contrast to this, the proposed method shows an average performance of 3.81 bpm across the three lighting conditions, as mentioned earlier, representing an improvement of 0.24 bpm MAE.

The performance on the 'talking' activity is worse as compared to that on other scene conditions for all three methods. Similar to other trends, the SNR weighting method shows a performance reduction of 0.05 bpm over the facial aggregation benchmark. However, the proposed method shows a large improvement of 0.57 bpm when compared to the facial aggregation benchmark. This large performance gain for the proposed method is further reinforced by looking at the correlation coefficient, which improves from 0.60 for the facial aggregation benchmark to 0.72 for the proposed method, as compared to 0.61 for the SNR weighting method.

Figure 4 highlights the high correlation between the proposed method's r-PPG HR estimates and ground truth PPG HR across the various recording conditions. The dark skin tone markers across all recording conditions make up the majority of outlying data. The B&A plots in Figures 4e-g show a bias of less than 1 bpm across the three lighting conditions, and Figure 4h shows a bias of less than 2 bpm during subject talking. These figures also show that the proposed method's r-PPG heart estimates are mostly within 10 bpm of the ground truth across all recording conditions. These correlation metrics are an improvement to the benchmark methods of facial aggregation and SNR weighting. Supplementary Fig. 3 and Supplementary Fig. 4 show the corresponding scatter and B&A plots for the facial aggregation and SNR weighting methods respectively.

**Camera viewpoint performance**

The final scene analysis factor is camera viewpoint. The VITAL dataset consists of two camera angles: *front*, where the camera is in front of the face, aimed directly at it, and *bottom*, where the camera is in front of the face but at a dip, so that the camera looks up to the face.

The facial aggregation method shows a MAE of 5.24 bpm for the *front* setting, and 3.74 bpm for the *bottom* setting. The SNR weighting method on the other hand shows MAEs of 5.38 bpm and 4.24 bpm. This represents a decrease in performance of 0.14 bpm and 0.50 bpm respectively. In contrast to this, the proposed method shows performance improvements of 0.35 bpm and 0.30 bpm MAE.

Supplementary Fig. 5, Supplementary Fig. 6, and Supplementary Fig. 7 show the corresponding scatter and B&A plots for the proposed method, facial aggregation method and SNR weighting method respectively. The correlation between estimated and ground truth HR seen by the proposed method for the *front* and *bottom* viewpoints (0.75 and 0.83) is a clear improvement over the same for the facial aggregation (0.68 and 0.80) and the SNR weighting (0.66 and 0.74) methods. The B&A plots also highlight the fact that the proposed method sees reduced bias, as well as reduced spread of HR for both the viewpoints, when viewed across skin tones as well as the recording conditions.

**DISCUSSION**

In this paper, we propose a novel r-PPG algorithm that mitigates performance losses for subjects with darker skin tones, subjects in varying illumination conditions, or subjects who may be moving their face such as when they are talking. The proposed method achieves the best overall average performance across the VITAL dataset of 4.17 bpm MAE, as opposed to 4.49 bpm by the facial aggregation method [5,7,10,13,25,26] and 4.81 bpm for the SNR weighting method [8,9,11,32]. Reinforcing this observation, similar improvement trends are observed for the proposed method, in terms of r value and SE over the facial aggregation benchmark.

We highlight the performance reduction of the SNR weighting method, and attribute it to its biased performance on lighter skin tones. Across all three skin tones, the SNR weighting method shows performance gain only for the light skin tone subjects and a performance drop for the other two skin tones, thereby actually increasing the skin tone performance bias. On the other hand, the proposed method shows increasing improvements in MAE across light, medium and dark skin tone. In fact, the proposed method shows both the best MAE performance in comparison with the facial aggregation and SNR weighting methods. This illustrates the importance for the need of a truly diverse dataset when developing r-PPG technology. Previous works that show overall performance enhancements on datasets with subjects of predominantly one skin tone may in fact not see the same results when applying their algorithms on more diverse datasets, such as the VITAL dataset.

The proposed method also achieves the largest MAE improvements of 0.55 bpm for the traditionally worse performing dark skin tone in comparison with the light and medium skin tones. This outcome attests to the fairness of the method. The proposed method is the only method able to perform with a MAE being less than 6 bpm across all skin tones. These inferences are further enforced by the largest increase in the correlation coefficient and largest decrease in the SE for dark skin tones by the proposed method, as opposed to the SNR weighting method which sees performance reduction for medium and dark skin tones. Hence, in addition to the overall improvement in performance across all skin tones, the proposed method successfully steps towards reducing the performance bias that exists between skin tones. If the VITAL dataset were to have even more equal representation in terms of skin tone, the overall average performance measures are further expected to improve. These findings are promising for the development and clinical adoption of high quality medically inclusive HR sensing technology that can be deployed on a smartphone and provide accurate HR estimates during a telemedicine visit.

Large improvements in performance of the proposed method are also observed for the talking activity over the facial aggregation benchmark, as compared to the SNR weighting method which shows an overall performance drop. As HR estimation with the proposed method provides improved performance for the talking setting, this technology may one day allow for real-time continuous contact-less HR monitoring during a telemedicine visit, which would provide greater information to outpatient clinicians. This advance may also be relevant for in-hospital continuous contactless monitoring in ICU settings or hospital floor care.

Improvements in performance are also observed across camera viewpoints. The proposed method shows considerable improvements for the *front* and *bottom* angles. This compares to performance drops for the SNR weighting method. A typical telemedicine visit, through a cell phone platform, may involve the patient holding the camera at varying angles with respect to the face. The proven robustness as well as performance improvement brought about by the proposed method therefore makes it increasingly amenable to such tasks, thereby bringing remote HR estimation algorithms closer to such deployments. Interestingly, for all methods tested (existing and novel), the *bottom* angle shows improved performance as compared to the *top* angle. This could be because interfering factors such as hair, spectacles and so on occupy a smaller portion of the usable frame in the *bottom* angle.

Beyond the algorithmic innovation of this work, another key contribution is the creation of the VITAL dataset, which is a first effort towards collecting a demographically diverse video vital sign database for telemedicine applications. While societal demographics are skewed largely towards light skin tone persons, it is essential to have diversely represented computer vision healthcare datasets, like the VITAL dataset, in order to explicitly understand performance limitations that may otherwise be masked within biased data [48]. While the VITAL dataset itself is not entirely unbiased itself, it achieves a much higher degree of skin tone diversity as compared to existing datasets. We therefore envision it to be an essential resource for upcoming related research and, in addition, to set the tone for future data collection endeavors for similar interdisciplinary clinical cum technological applications.

In relation to the significance of this work, remote vital sign monitoring has risen in prominence over recent years, with an acceleration in clinical development due to the COVID-19 pandemic. We envision that the development and validation of computer vision algorithms that facilitate non-contact vital sign sensing will have implications for telemedicine, remote patient monitoring, and in-person care. In response to the pandemic, health systems across the country implemented a large-scale restriction of non-urgent in-person appointments [49], transitioned many outpatient services to telemedicine visits [3], and developed remote monitoring care pathways [1] in order to facilitate social distancing yet maintain continuity of care. To remotely monitor COVID-19 patients, many health systems shipped home vital sign equipment to patients in order to obtain quantitative physiological data that could facilitate high quality remote management via telemedicine. At a population level, however, supplying and shipping vital sign monitoring devices to patients is expensive and not scalable, making such a solution nonviable, especially during a global pandemic. Given the high penetration of mobile phone technology globally, there is great interest in transforming smartphones into low-cost portable HR, respiratory rate, and pulse oximeter monitors, with the intent of increasing accessibility to vital monitoring equipment in order to alleviate healthcare inequity. Smartphones that could use computer vision algorithms to obtain quantitative vital sign data in a contactless and remote manner would greatly improve the quality of telemedicine care, allowing clinicians to make remote medical decisions with quantitative physiological data and qualitative patient history. Outside of a pandemic situation, knowledge of vital signs is also important information for clinicians who are managing medical conditions that require such data for health management, and remotely obtaining vital signs may allow care teams to perform remote surveillance and home monitoring of patients with greater confidence. Notably, several patient populations may benefit from more remote care. It has been established that the COVID-19 pandemic has disproportionately affected African American and minority communities and those suffering from lower socioeconomic status in the United States, both nationally and in states the most affected by the pandemic [50,51]. In New York City and Michigan, African American and Latino residents have the highest age-adjusted rates of hospitalized and non-hospitalized COVID-19, and age-adjusted death rates for African Americans are more than twice those for white and Asian residents [52,53]. African American communities have also been found to have higher prevalence of cardiovascular and related complications, when compared with traditionally light skin toned people [59]. These patient populations may therefore stand to benefit the most from skin tone robust contactless vital sign (specifically heart rate) sensing technologies that facilitate high-quality remote care pathways, thereby reducing their exposure to healthcare settings or waiting areas that may have other infectious sources. Finally, we believe contactless vital sign sensing technology would be useful at the start of in-person clinic or hospital encounters or for continuous patient monitoring in a hospital floor or ICU setting. Cameras, as opposed to hospital staff, may one day obtain key vital signs without contact, thereby reducing exposure of patients to staff, enabling improved infection control, and freeing up hospital staff to attend to other important patient care needs.

With regards to limitations and future work, while our method has been tested on an adult population, additional work is needed to enable clinical adoption. Further research investigating HR estimation using our proposed method is still needed in pediatric and geriatric populations and patient populations with known cardiopulmonary disease. Future work must also focus on improving computer vision methods to detect extremes of HR and discern heart arrhythmias. Research must be undertaken to further improve overall performance on subjects and videos in real life scenarios and to continue to reduce bias and assure fairness by building upon our work.

Finally, from an algorithmic perspective, we believe that one of the most important factors towards large scale deployment of such methods for clinical use is the inherent fairness of the algorithm. As healthcare increasingly accelerates towards a digitally connected and virtual future, early consideration must be given to developing equitable health technology that does not exacerbate healthcare disparities or create new disparities. Ultimately, we hope this work motivates the computer vision community towards exciting and essential research avenues looking into inherent system biases associated with r-PPG. By reducing biases, we move a step closer towards deploying high-quality, medically inclusive non-contact vital sensing techniques that can aid clinicians in delivering remote patient care, during times of peace and pandemic alike.

## METHODS

### Data Collection Protocol

The human study protocol was approved by the UCLA Institutional Review Board (IRB#20-001025-AM-00001), and participants provided written informed consent to take part in the study. Figure 1 shows the data collection setup. Each subject is made to sit on a height-adjustable chair, in the field of view of two cell-phone cameras (with different view angles): one camera (Samsung Galaxy S10) is perfectly front-on, while the other (Samsung Galaxy A51) is directly in front of the face, at a dip (lower) of 15 degrees. The front-on camera is placed approximately 130 cm from the subject, and the lower camera at a dip is approximately 90 cm from the subject. The height of the chair is chosen so that the subject is centered in the front-on frame. The controlled lights are set up on either side of the front-on camera, with a baseline of 100 centimeters between them.

We record subjects using these cameras under four different scene conditions: (1) controlled lighting at 5600K ("cool" lighting) with the subject remaining stationary, (2) controlled lighting at 3200K ("warm" lighting) with the subject remaining stationary, (3) ambient room lighting- with distributed white LED lighting- with the subject remaining stationary, and (4) ambient room lighting with the subject speaking. Controlled lighting is enabled by a pair of professional bi-color LED photography lights. The controlled lighting recording conditions were performed with the room lights kept off, allowing for fine-tuned control over the illumination spectral properties. As incorporating controlled lighting only enables a front-facing illumination angle, two recording conditions in ambient room lighting were captured where the subject was lit more completely from several angles. The final recording condition involved variations in the subject, including talking, natural head movements, and facial expressions. Each scene recording session lasts for 2 minutes, for a total of 16 minutes of video footage across 8 videos.

During data collection, volunteers are fitted with standard anesthesiology cardiopulmonary monitors: pulse oximeter (Red DCI, Masimo), blood pressure cuff (Comfort Care, Philips), and 5-lead electrocardiogram (Philips IntelliVue). To collect vital sign data, we utilize the Philips IntelliVue MX800 patient monitor to perform real time monitoring of four vital signs- HR, respiratory rate, oxygen saturation, and non-invasive continuous blood pressure- of which three waveforms are collected (ECG, PPG and respiration). We use the open source tool *VSCapture* [54] to collect data onto a computer using the MX800's local area network communication protocol. The MX800's estimated numeric values for the vital signs are sampled every 1 second, while the waveforms are sampled at variable frequencies. The ECG signal is sampled between 400-600 Hz, the PPG signal between 100-150Hz and the respiration between 40-60Hz. Continuous non-invasive blood pressure estimates occur when the blood pressure cuff is activated, which is approximately once every 30 seconds.

A total of 60 subjects participated in the study. Due to data collection errors or corrupted video, 6 subjects are excluded from the experiment. Therefore, the final VITAL dataset consists of 432 videos (~864 minutes) of 54 subjects and their vital signs.

**Statistical analysis**

In order to quantitatively assess the performance of the proposed method, the following statistical metrics are used to: (i) Mean Absolute Error (MAE), (ii) Standard deviation of the error (SE) and the correlation coefficient (r) between the estimated r-PPG average HR and the ground truth PPG average HR for the entire video. We also employ Bland-Altman (B&A) plots [55] to compare differences in the proposed method's HR estimates and MX800 PPG HR measurements (Figures 3 and 4). These plots are labelled with the corresponding mean difference (m) that shows the systematic bias, and the limits of agreement (LoA) within which 95% of the differences are expected to lie, estimated as LoA = m ± 1.96 σ, assuming a normal distribution.

**Bias in r-PPG**

While r-PPG based HR estimation has been increasingly researched over the past years, certain inherent biases and performance gaps continue to exist, across subject and scene conditions. These biases include: (i) subject skin tone, (ii) scene lighting, (iii) shadows and specular highlights (bright regions in an image which are reflections of the light source, rather than transmissions from the skin) and (iv) facial motion due to talking.   In what follows, we use first principles to derive potential sources of bias, link biases to statistical noise, and develop novel denoising and debiasing algorithms, whose source code is available.

**Effect of Skin Tone on PPG Signal**

The goal of this subsection is to use light transport theory to show that appreciable error in r-PPG estimation due to dark skin is not due to biophysical factors, but instead due to imaging noise. Previous work has developed a mathematical model for skin coloration, as a function of melanin content and blood volume fraction [56]. We extend this existing coloration model for the new goal of analyzing response of the PPG signal to interference and noise ratio (SINR), as well as PPG signal strength, in the context of skin tone variation. Let $E(\lambda)$ represent the spectral power distribution of the light source concerned. Let $S_c(\lambda)$ be the spectral sensitivity of the camera in use for color channel $c$. The model we follow assumes that light from the skin, as seen by the camera, emerges after two transmissions from the epidermis and one reflection from the dermis. That is, $R(\lambda) = T^2{}_{ep}(\lambda).R_d(\lambda)$. Using the expressions for $T(\lambda)$ and $R_d(\lambda)$ derived in Alotaibi et al. [56], we can evaluate the value of $R(\lambda)$, as a parametric function of $f_{mel}$ (skin melanin fraction), $f_{blood}$ (fraction of blood in the specific skin region) and $f_{hg}$ (fraction of hemoglobin in the blood at the location). Then, the intensity captured in channel $c$ by the camera is given by $\int_\lambda E(\lambda)S_c(\lambda)R(\lambda)d\lambda$. Subsequently, we refer to $R(\lambda)$ as $R(\lambda, f_{mel}, f_{blood}, f_{hg})$ to incorporate all the relevant parameters. To understand the SINR as a function of radiance frequency, we identify that the PPG signal arises out of temporal variation in the value of $f_{blood}$. On the other hand, the interference consists of the average skin tone value, on which the visual PPG signal rides. The noise involved is the noise involved in the capture process through the camera. First, we look at the Signal to Interference Ratio (SIR) and signal strength, while ignoring the effect of imaging noise (analyzed in the next subsection). The PPG signal strength may be approximated by

$$S(\lambda) = \frac{dR}{df_{blood}} \Delta f_{blood}.$$

Since the variation in $f_{blood}$ induces a small change in the skin color visible to the camera, the above approximation is valid. Similarly, the interference strength, which is the average skin tone value may be approximated by $R(\lambda, f_{mel}, f_{blood}, f_{hg})|_{f_{blood}=\underline{f_{blood}}}$, where $\underline{f_{blood}}$ is the average blood volume fraction. The SIR for a given frequency is given by,

$$L(\lambda) = \frac{|\frac{dR}{df_{blood}}\Delta f_{blood}|^2}{|R(\lambda, f_{mel}, f_{blood}, f_{hg})|_{f_{blood}=\underline{f_{blood}}}|^2}.$$

Then, an estimate for the average signal strength is given by $M = \int_\lambda E(\lambda)S_c(\lambda)S(\lambda)d\lambda$, and the SIR across frequencies is given by $N = \int_\lambda E(\lambda)S_c(\lambda)L(\lambda)d\lambda$. These metrics $M$ and $N$ are parametrized by $f_{mel}$ and $f_{hg}$. Since we are interested in analyzing the effect of skin tone, we hold $f_{hg}$ constant and evaluate the above SIR metric for various reasonable values of $f_{mel}$. The values of other relevant physiological constants are taken to be as defined in Alotaibi et. al [56], i.e. taken to be the average healthy values. It can be seen in the expression for $L(\lambda)$ that the melanin dependent term cancels out between the numerator and denominator. Hence, the (biophysical) SIR is independent of the skin melanin content.

Figure 5a. shows the signal strength curves $S(\lambda)$ for different skin melanin fractions. As is clear, as well as intuitively understood, the signal strength reduces with increasing skin melanin content. We can therefore infer that the corruption added to the signal, that hinders accurate inference, is not biophysical in nature (observed from the constant biophysical SIR). In contrast, the decreasing signal strength leads us to an analysis of imaging noise, which is the major noise phenomenon at play in this case.

**Effect of Imaging Noise on PPG Signal and Algorithms**

Having analyzed the SIR in the previous section, we now look at SNR trends. The goal of this subsection is to understand the relationship between imaging noise and r-PPG algorithm estimation. Imaging noise refers to the inherent noise that arises due to the image capture process in a commercial camera. This arises due to various effects related to photon arrival processes, thermal noise in electronics and the quantization noise associated with digitally capturing images [57]. Overall, the entire signal to noise ratio for a pixel of a particular intensity is given by:

$$S(\lambda) = \frac{p}{\sqrt{(\frac{p}{g} + (\frac{\sigma_r}{g})^2 + (\sigma_q)^2)}},$$

where $p$ is the pixel value (ranging from 0-255), $g$ is the sensor gain (a constant for a given image) and $\sigma_r$ and $\sigma_q$ are camera noise parameters (also constant). Plugging in typical values for the constants, Figure 5b. shows the trend for the SNR as a function of pixel value. The SNR is smaller for lower pixel values (corresponding to darker skin or shadowed regions) as compared to higher pixel values (corresponding to brighter skin or lit up regions). These observations, coupled with the observations from the previous subsection, allow us to make the following inferences:

(i) *Imaging noise creates skin tone bias (and lighting bias):* The performance gap across skin tones, as well as across lighting differences, can be understood in terms of imaging noise. Darker skin regions have lower signal strength that manifest as lower pixel values in the video. This results in poorer SNRs. Note that this inference also holds true for shadowed regions, thereby extending this analysis towards understanding lighting bias.

(ii) *Imaging noise and specular reflections degrade the r-PPG signal:* Since the biophysical SIR is independent of skin tone, the imaging noise, coupled with specular highlights due to lighting, are the major contributing factors to signal degradation. The corruption due to imaging noise depends on signal intensity, as described above. The corruption due to specular highlights depends on lighting conditions- regions with strong specular highlights have relatively lower PPG signal information. Combating the highlighted biases in existing r-PPG would therefore involve a principled approach towards reduction of the above highlighted imaging noise and specular highlight removal. Note that specular highlight removal, in addition to reducing lighting related biases, also indirectly affects skin tone bias: darker skin subjects are worse affected by these interferences, since the intensity difference between the signal and the highlight is much more for them.

(iii) *Denoising to be done before signal inference:* This noise removal must be carried out in the combination step (defined below) as opposed to after signal aggregation.

With the above inferences on hand, we look at the performance of existing methods that introduce averaging techniques in the combination step to improve the signal to noise ratio.

**The r-PPG Pipeline**

We first describe in detail a typical remote PPG algorithm pipeline for ease of understanding for the reader. There are four distinct components to this pipeline: (a) detection, which identifies facial regions of interest in the video frame, (b) combination, which condenses the information from regions of interest into a RGB time series signal , (c) signal inference, which uses the time series signal to estimate the pulse volume waveform, and (d) HR estimation, which estimates the HR from the pulse volume signal.  This pipeline is visually described in Figure 6.

The video is first passed through a neural network based face detector [46], in order to identify the face region in the frame. Using feature point detectors [47], the eye and mouth regions are identified and explicitly removed from the videos (since these regions do not contribute to the pulsatile signal). This is the *detection step*. The next steps, namely combination, inference and HR step, are carried out for smaller video-windows of 10 seconds length with an overlap of 5 seconds.

For each video frame, now, the skin pixels are combined together to get one RGB sample for that time instance (the methods for this combination vary across papers and is the crux of this work's novelty). Across all frames, after this combination, we obtain a time series RGB signal. This is the *combination step*.

These RGB signals are then put through an existing signal inference technique. In this paper, we use the CHROM algorithm [7] due to its versatility, as well as its easy access from openly available code [45]. The output obtained from this step results in a pulsatile waveform estimate for each window. This is the *inference step*.

The obtained pulsatile waveform is then processed to arrive at the final HR. This is the *heart rate step*. We first filter the waveform using a 3rd-order Butterworth bandpass filter with pass band frequencies of [0.7, 3.5] Hz. The power spectral density (PSD) is then computed. Temporal frequency artifacts were empirically observed in the original video as a result of aggressive compression, likely due to the unchanging green background. These erroneous peaks were appropriately removed. Next, the five highest peaks in the PSD are chosen. The peak with the highest combined fundamental and second harmonic power is chosen as the one corresponding to the HR. The final HR for the video is estimated as the average of the HR estimates for each 10 second window.

**Performance of Existing Methods**

We look at existing algorithms that propose methods to improve the noise performance in the combination step. The most straightforward approach is to simply average all face pixels in a frame, in order to arrive at time samples of the RGB signal. We refer to this as facial aggregation [5,7,10,13,25,26]. To improve upon this, previous approaches have sought to modify this averaging process. We describe the best performing result amongst these on the VITAL dataset. The face is gridded into smaller rectangular regions. Pixels within each region are averaged to arrive at individual time series for each region. Each of these gridded temporal signals is passed through the inference step, to obtain the corresponding blood volume signal estimate. Approaches use measures such as SNR at peak frequency of this blood volume signal to characterize the 'goodness' of each signal [7–9,11,32], with higher weights being assigned for better signals. As mentioned previously, in this paper we use the two harmonic SNR estimate, which was found to be more robust. That is, for a signal $s$ (frequency domain $S$) with a HR $p$, the SNR at the HR frequency is given by:

$$SNR = \frac{\int_{p-w}^{p+w} |S(f)|^2 df + \int_{2p-2w}^{2p+2w} |S(f)|^2 df}{\int_{-\infty}^{\infty} |S(f)|^2 df - \int_{p-w}^{p+w} |S(f)|^2 df - \int_{2p-2w}^{2p+2w} |S(f)|^2 df},$$

where $w$ is the peak window size for estimation (for this work's experiments, we use $w$ =0.1 Hz).

This resultant signal is passed to the HR step. We call this method SNR weighting [8,9,11,32]. Finally, these weights are used to average the blood volume signals together. Table 2 and table 3 shows the results for facial aggregation and SNR weighting. As can be seen, the method affords an improvement for light skin tones, but shows a stark reduction in performance for medium and darker skin tones. This performance reduction can be understood in terms of the weight maps. The weight maps from previous methods (based on region-based SNR estimates) have the tendency to be sparse, especially for darker skin tones. The improvements due to weighted averaging are therefore lost to noise corruption for darker skin tone subjects since much lesser signal is being aggregated. As per our analysis, this poorer denoising for darker skin subjects results in worse SNRs, thereby degrading performance. Datasets on which these previous methods were tested were not as diverse across skin tones: these performance caveats were therefore missed.

Additionally, the previous method of SNR weighting may also fall prey to specular highlights. As described earlier, specular highlights are regions of the face where the light from the source is directly reflected from the skin to the camera. While the signal intensity (pixel value) is high, the signal contains no information of the pulsatile signal, which gets buried in the light from the source. This is a considerable factor when looking at scene conditions, such as camera angle, lighting direction, lighting color and intensity, as well as skin tone (since specular highlights affect darker skin more than lighter skin). Previous weighting approaches do not explicitly take this into account and use the gridded weighting method to implicitly combat these highlights. However, since the nature of this gridding itself degrades for darker skin tones, we observe that specular effects must be directly addressed.

**Novel Modifications**

Having identified the reasons for poor performance of existing methods, we propose novelties to be incorporated in the combination step, that look to achieve a performance gain in a manner that is fair across skin tones. Specifically, there are two major novelties that we propose: (i) weighting in RGB space rather than blood volume signal space and (ii) skin diffuse component weighting. We now describe each of these steps in detail.

(a) RGB-space weighting: Existing spatial averaging methods estimate weights for each grid region, based on the blood volume signal quality [8,9,11,32]. Instead of using these estimated weights to average the blood volume signals, as done in previous methods, we propose using these weights to average in RGB space. As a result, we obtain one consolidated SNR weighted RGB signal, which is again passed through the inference step to obtain the final blood volume signal.

The motivation for this modification can be understood in the context of noise. Averaging the RGB signal before passing through the inference step results in a less noisy signal passing through the inference method, enabling the inference method to provide better estimates, as compared to when noisier signals are passed through the method, to be averaged later. If the inference method is non-linear (such as CHROM), a pre weighting would lead to additional noise performance gain.

(b) Skin diffuse component weighting: An image can be split into two constituent components: the diffuse component, that arises out of transmission and reflection through the skin, and a specular component, that arises from mirror-like surface reflections. Since the diffuse component contains the signal of interest for us, we propose, for the first time, the usage of gridded diffuse components as additional weights. For each frame, the diffuse component is estimated [58]. It is then gridded and averaged across the grid dimensions and time, in order to arrive at weights for each grid element.

The diffuse weights play two key roles in improving bias in performance as well as overall performance: first, they can remove specular affected regions from the average explicitly. Second, they combat the sparsity issue observed in traditional SNR weights, since the diffuse component is continuous and non-sparse. The SNR weights and the novel diffuse weights are multiplied together and renormalized to arrive at the final spatial weights for the gridded video. The overall pipeline, therefore, involves using the novel weights together, to arrive at efficiently weighted RGB signals. These are averaged together and passed through the estimation step and HR step. This pipeline is visually highlighted as such in Figure 6.

**Data Availability**

The data that has been used to support the findings of this study is available from the corresponding author upon request and adherence to IRB protocols.

**Code Availability**

The code/software that has been used to support the findings of this study is available from the corresponding author upon request and adherence to IRB protocols.

**Acknowledgements**

The work of UCLA Department of Electrical and Computer Engineering authors is supported by a Sony Imaging Young Faculty Award, Google Faculty Award, and the NSF CRII Research Initiation Award (IIS 1849941). M.C. is supported by the NIH under award numbers R01HL144692 and R01EB029751.

**Competing Interests**

M.C. has ownership interest in Sironis and Perceptive Medical, companies developing closed-loop systems. M.C. is consulting for Edwards Lifesciences (Irvine, CA) and Masimo Corp. (Irvine, CA). M.C. has received research support from Edwards Lifesciences through his Department. A.K. has ownership interest in Akasha Imaging, a computer vision company working on robotic imaging problems.

**Author Contributions**

P.C., K. Kabra, and A.K. conceptualized the overall design of the algorithm. P.C., D.S. and T.C. worked on the detection and combination steps of the proposed algorithm. D.K. and K. Kulkarni worked on the HR estimation step. D.K. implemented comparison benchmarks. P.C. and A.K. conceptualized the theory, which P.C derived and simulated. L.J. and A.K. initiated the IRB for data collection. P.C., K. Kabra and L.J. worked on organizing the collection and storage of data. P.C., K. Kabra, S.L., M.C., L.J., and A.K. wrote the manuscript. M.C., L.J, and A.K. conceptualized the study. A.K. oversaw the project.

**REFERENCES**

1. Annis, T. *et al.* Rapid implementation of a COVID-19 remote patient monitoring program. *J. Am. Med. Inform. Assoc.* **27**, 1326–1330 (2020).
2. Ford, D. *et al.* Leveraging health system telehealth and informatics infrastructure to create a continuum of services for COVID-19 screening, testing, and treatment. *J. Am. Med. Inform. Assoc.* doi:10.1093/jamia/ocaa157.

3. Connolly, S. L. *et al.* Rapid Increase in Telemental Health Within the Department of Veterans Affairs During the COVID-19 Pandemic. *Telemed. E-Health* (2020) doi:10.1089/tmj.2020.0233.

4. Nishiga, M., Wang, D. W., Han, Y., Lewis, D. B. & Wu, J. C. COVID-19 and cardiovascular disease: from basic mechanisms to clinical perspectives. *Nat. Rev. Cardiol.* **17**, 543–558 (2020).

5. Poh, M.-Z., McDuff, D. J. & Picard, R. W. Non-contact, automated cardiac pulse measurements using video imaging and blind source separation. *Opt. Express* **18**, 10762–10774 (2010).

6. Balakrishnan, G., Durand, F. & Guttag, J. Detecting Pulse from Head Motions in Video. in *2013 IEEE Conference on Computer Vision and Pattern Recognition* 3430–3437 (2013). doi:10.1109/CVPR.2013.440.

7. de Haan, G. & Jeanne, V. Robust Pulse Rate From Chrominance-Based rPPG. *IEEE Trans. Biomed. Eng.* **60**, 2878–2886 (2013).

8. Po, L.-M. *et al.* Block-based adaptive ROI for remote photoplethysmography. *Multimed. Tools Appl.* **77**, 6503–6529 (2018).

9. Li, P., Benezeth, Y., Nakamura, K., Gomez, R. & Yang, F. Model-based Region of Interest Segmentation for Remote Photoplethysmography. in *Proceedings of the 14th International Joint Conference on Computer Vision, Imaging and Computer Graphics Theory and Applications (VISAPP)* vol. 4 383–388 (2020).

10. Wang, W., Stuijk, S. & Haan, G. de. A Novel Algorithm for Remote Photoplethysmography: Spatial Subspace Rotation. *IEEE Trans. Biomed. Eng.* **63**, 1974–1984 (2016).

11. Kumar, M., Veeraraghavan, A. & Sabharwal, A. DistancePPG: Robust non-contact vital signs monitoring using a camera. *Biomed. Opt. Express* **6**, 1565–1588 (2015).

12. Moço, A. V., Stuijk, S. & Haan, G. de. Motion robust PPG-imaging through color channel mapping. *Biomed. Opt. Express* **7**, 1737–1754 (2016).

13. Wang, W., den Brinker, A. C., Stuijk, S. & de Haan, G. Algorithmic Principles of Remote PPG. *IEEE Trans. Biomed. Eng.* **64**, 1479–1491 (2017).

14. Addison, P. S., Jacquel, D., Foo, D. M. H., Antunes, A. & Borg, U. R. Video-Based Physiologic Monitoring During an Acute Hypoxic Challenge: Heart Rate, Respiratory Rate, and Oxygen Saturation. *Anesth. Analg.* **125**, 860–873 (2017).

15. Cobos-Torres, J.-C., Abderrahim, M. & Martínez-Orgado, J. Non-Contact, Simple Neonatal Monitoring by Photoplethysmography. *Sensors* **18**, 4362 (2018).

16. Villarroel, M. *et al.* Non-contact physiological monitoring of preterm infants in the Neonatal Intensive Care Unit. *Npj Digit. Med.* **2**, 1–18 (2019).

17. Chen, W. & McDuff, D. DeepPhys: Video-Based Physiological Measurement Using Convolutional Attention Networks. in *Computer Vision – ECCV 2018* (eds. Ferrari, V., Hebert, M., Sminchisescu, C. & Weiss, Y.) 356–373 (Springer International Publishing, 2018). doi:10.1007/978-3-030-01216-8_22.

18. Niu, X., Shan, S., Han, H. & Chen, X. RhythmNet: End-to-End Heart Rate Estimation From Face via Spatial-Temporal Representation. *IEEE Trans. Image Process.* **29**, 2409–2423 (2020).

19. Yu, Z., Peng, W., Li, X., Hong, X. & Zhao, G. Remote Heart Rate Measurement From Highly Compressed Facial Videos: An End-to-End Deep Learning Solution With Video Enhancement. in *2019 IEEE/CVF International Conference on Computer Vision (ICCV)* 151–160 (2019). doi:10.1109/ICCV.2019.00024.

20. Aubakir, B., Nurimbetov, B., Tursynbek, I. & Varol, H. A. Vital sign monitoring utilizing Eulerian video magnification and thermography. in *2016 38th Annual International Conference of the IEEE Engineering in Medicine and Biology Society (EMBC)* 3527–3530 (2016). doi:10.1109/EMBC.2016.7591489.

21. Song, R., Zhang, S., Cheng, J., Li, C. & Chen, X. New insights on super-high resolution for video-based heart rate estimation with a semi-blind source separation method. *Comput. Biol. Med.* **116**, 103535 (2020).

22. Tulyakov, S. *et al.* Self-Adaptive Matrix Completion for Heart Rate Estimation from Face Videos under Realistic Conditions. in *2016 IEEE Conference on Computer Vision and Pattern Recognition (CVPR)* 2396–2404 (2016). doi:10.1109/CVPR.2016.263.

23. Yu, Z., Li, X. & Zhao, G. Remote Photoplethysmograph Signal Measurement from Facial Videos Using Spatio-Temporal Networks. Preprint at https://arxiv.org/abs/1905.02419 (2019).

24. Tsouri, G. R., Kyal, S., Dianat, S. A. & Mestha, L. K. Constrained independent component analysis approach to nonobtrusive pulse rate measurements. *J. Biomed. Opt.* **17**, 077011 (2012).

25. Lewandowska, M., Rumiński, J., Kocejko, T. & Nowak, J. Measuring pulse rate with a webcam — A non-contact method for evaluating cardiac activity. in *2011 Federated Conference on Computer Science and Information Systems (FedCSIS)* 405–410 (2011).

26. de Haan, G. & van Leest, A. Improved motion robustness of remote-PPG by using the blood volume pulse signature. *Physiol. Meas.* **35**, 1913–1926 (2014).

27. Nowara, E., McDuff, D. & Veeraraghavan, A. The Benefit of Distraction: Denoising Remote Vitals Measurements using Inverse Attention. Preprint at https://arxiv.org/abs/2010.07770v1 (2020).

28. Nowara, E. M., Marks, T. K., Mansour, H. & Veeraraghavan, A. SparsePPG: Towards Driver Monitoring Using Camera-Based Vital Signs Estimation in Near-Infrared. in *2018 IEEE/CVF Conference on Computer Vision and Pattern Recognition Workshops (CVPRW)* 1353–135309 (2018). doi:10.1109/CVPRW.2018.00174.

29. Li, X., Chen, J., Zhao, G. & Pietikäinen, M. Remote Heart Rate Measurement from Face Videos under Realistic Situations. in *2014 IEEE Conference on Computer Vision and Pattern Recognition* 4264–4271 (2014). doi:10.1109/CVPR.2014.543.

30. Wang, W., Stuijk, S. & Haan, G. de. Exploiting Spatial Redundancy of Image Sensor for Motion Robust rPPG. *IEEE Trans. Biomed. Eng.* **62**, 415–425 (2015).

31. Verkruysse, W., Svaasand, L. O. & Nelson, J. S. Remote plethysmographic imaging using ambient light. *Opt. Express* **16**, 21434–21445 (2008).

32. Bobbia, S., Macwan, R., Benezeth, Y., Mansouri, A. & Dubois, J. Unsupervised skin tissue segmentation for remote photoplethysmography. *Pattern Recognit. Lett.* **124**, 82–90 (2019).

33. Tang, C., Lu, J. & Liu, J. Non-contact Heart Rate Monitoring by Combining Convolutional Neural Network Skin Detection and Remote Photoplethysmography via a Low-Cost Camera. in *2018 IEEE/CVF Conference on Computer Vision and Pattern Recognition Workshops (CVPRW)* 1390–13906 (2018). doi:10.1109/CVPRW.2018.00178.

34. Nagamatsu, G., Nowara, E. M., Pai, A., Veeraraghavan, A. & Kawasaki, H. PPG3D: Does 3D head tracking improve camera-based PPG estimation? in *2020 42nd Annual International Conference of the IEEE Engineering in Medicine Biology Society (EMBC)* 1194–1197 (2020). doi:10.1109/EMBC44109.2020.9176065.

35. Spetlík, R., Franc, V., Cech, J. & Matas, J. Visual Heart Rate Estimation with Convolutional Neural Network. in *BMVC* (2018).

36. Poh, M., McDuff, D. J. & Picard, R. W. Advancements in Noncontact, Multiparameter Physiological Measurements Using a Webcam. *IEEE Trans. Biomed. Eng.* **58**, 7–11 (2011).

37. Sun, G. *et al.* Remote sensing of multiple vital signs using a CMOS camera-equipped infrared thermography system and its clinical application in rapidly screening patients with suspected infectious diseases. *Int. J. Infect. Dis. IJID Off. Publ. Int. Soc. Infect. Dis.* **55**, 113–117 (2017).

38. Patil, O., Wang, W., Gao, Y. & Jin, Z. MobiEye: turning your smartphones into a ubiquitous unobtrusive vital sign monitoring system. *CCF Trans. Pervasive Comput. Interact.* **2**, 97–112 (2020).

39. Nowara, E. M., McDuff, D. & Veeraraghavan, A. A Meta-Analysis of the Impact of Skin Tone and Gender on Non-Contact Photoplethysmography Measurements. in 284–285 (2020).

40. Mosa, A. S. M., Yoo, I. & Sheets, L. A Systematic Review of Healthcare Applications for Smartphones. *BMC Med. Inform. Decis. Mak.* **12**, 67 (2012).

41. Ventola, C. L. Mobile Devices and Apps for Health Care Professionals: Uses and Benefits. *Pharm. Ther.* **39**, 356–364 (2014).

42. Boulos, M. N. K., Wheeler, S., Tavares, C. & Jones, R. How smartphones are changing the face of mobile and participatory healthcare: an overview, with example from eCAALYX. *Biomed. Eng. OnLine* **10**, 24 (2011).

43. Nowara, E. & McDuff, D. Combating the Impact of Video Compression on Non-Contact Vital Sign Measurement Using Supervised Learning. in *2019 IEEE/CVF International Conference on Computer Vision Workshop (ICCVW)* 1706–1712 (2019). doi:10.1109/ICCVW.2019.00211.

44. Fitzpatrick, T. B. The Validity and Practicality of Sun-Reactive Skin Types I Through VI. *Arch. Dermatol.* **124**, 869–871 (1988).

45. McDuff, D. & Blackford, E. iPhys: An Open Non-Contact Imaging-Based Physiological Measurement Toolbox. Preprint at https://arxiv.org/abs/1901.04366 (2019).

46. Zhang, K., Zhang, Z., Li, Z. & Qiao, Y. Joint Face Detection and Alignment Using Multitask Cascaded Convolutional Networks. *IEEE Signal Process. Lett.* **23**, 1499–1503 (2016).

47. Kazemi, V. & Sullivan, J. One millisecond face alignment with an ensemble of regression trees. in *2014 IEEE Conference on Computer Vision and Pattern Recognition* 1867–1874 (2014). doi:10.1109/CVPR.2014.241.

48. Cahan, E. M., Hernandez-Boussard, T., Thadaney-Israni, S. & Rubin, D. L. Putting the data before the algorithm in big data addressing personalized healthcare. *Npj Digit. Med.* **2**, 1–6 (2019).

49. Jm, F. *et al.* Virtual Care Expansion in the Veterans Health Administration During the COVID-19 Pandemic: Clinical Services and Patient Characteristics Associated with Utilization. *J. Am. Med. Inform. Assoc. JAMIA* (2020) doi:10.1093/jamia/ocaa284.

50. Abedi, V. *et al.* Racial, Economic, and Health Inequality and COVID-19 Infection in the United States. *J. Racial Ethn. Health Disparities* (2020) doi:10.1007/s40615-020-00833-4.

51. Azar, K. M. J. *et al.* Disparities In Outcomes Among COVID-19 Patients In A Large Health Care System In California. *Health Aff. (Millwood)* **39**, 1253–1262 (2020).

52. Holtgrave, D. R., Barranco, M. A., Tesoriero, J. M., Blog, D. S. & Rosenberg, E. S. Assessing racial and ethnic disparities using a COVID-19 outcomes continuum for New York State. *Ann. Epidemiol.* **48**, 9–14 (2020).

53. Gu, T. *et al.* Characteristics Associated With Racial/Ethnic Disparities in COVID-19 Outcomes in an Academic Health Care System. *JAMA Netw. Open* **3**, e2025197 (2020).

54. Karippacheril, J. G. & Ho, T. Y. Data acquisition from S/5 GE Datex anesthesia monitor using VSCapture: An open source.NET/Mono tool. *J. Anaesthesiol. Clin. Pharmacol.* **29**, 423–424 (2013).

55. Altman, D. G. & Bland, J. M. Measurement in Medicine: The Analysis of Method Comparison Studies. *J. R. Stat. Soc. Ser. Stat.* **32**, 307–317 (1983).

56. Alotaibi, S. & Smith, W. A. P. A Biophysical 3D Morphable Model of Face Appearance. in *2017 IEEE International Conference on Computer Vision Workshops (ICCVW)* 824–832 (2017). doi:10.1109/ICCVW.2017.102.

57. Hasinoff, S. W., Durand, F. & Freeman, W. T. Noise-optimal capture for high dynamic range photography. in *2010 IEEE Computer Society Conference on Computer Vision and Pattern Recognition* 553–560 (2010). doi:10.1109/CVPR.2010.5540167.

58. Yang, Q., Wang, S. & Ahuja, N. Real-Time Specular Highlight Removal Using Bilateral Filtering. in *Computer Vision – ECCV 2010* (eds. Daniilidis, K., Maragos, P. & Paragios, N.) 87–100 (Springer, 2010). doi:10.1007/978-3-642-15561-1_7.

59. Mensah G. A. (2018). Cardiovascular Diseases in African Americans: Fostering Community Partnerships to Stem the Tide. *American journal of kidney diseases : the official journal of the National Kidney Foundation*, *72*(5 Suppl 1), S37–S42. https://doi.org/10.1053/j.ajkd.2018.06.026

**TABLES**

| Table 1. Demographic characteristics of volunteers in the VITAL dataset. | | |
|---|---|---|
| Total number of participants in study | | 54 |
| **Physical Demographics** | **Mean** | **Median** |
| Age (years) | 34 (10) | 34 (26-41) |
| Height (cm) | 173 (9) | 175 (164-180) |
| Weight (kg) | 72 (16) | 72 (56-81) |
| Body Mass Index (kg m$^{-2}$) | 24 (5) | 23 (21-26) |
| | | |
| **Sex** | **# of participants** | |
| Male | 33 (61%) | |
| Female | 21 (39%) | |
| | | |
| **Race** | **# of participants** | |
| White | 27 (50%) | |
| Asian | 16 (29%) | |
| Black or African American | 8 (15%) | |
| Native Hawaiian or other Pacific Islander | 0 (0%) | |
| American Indian or Alaska Native | 2 (4%) | |
| Unknown | 1 (2%) | |
| | | |
| **Ethnicity** | **# of participants** | |
| Hispanic/Latino | 7 (13%) | |
| non-Hispanic/Latino | 47 (87%) | |
| | | |
| **Skin Type** | **# of participants** | |
| Light | 19 (35%) | |
| Medium | 24 (45%) | |
| Dark | 11 (20%) | |

| Pre-processing | Statistic | Skin Type | | | Recording Condition | | | | Camera viewpoint | | Overall |
|---|---|---|---|---|---|---|---|---|---|---|---|
| | | Light | Medium | Dark | 3200K | 5600K | Room Lighting | Talking | Front | Lower | |
| Facial aggregation | MAE (bpm) | 3.94 | 4.14 | 6.20 | 3.91 | 4.24 | 3.99 | 5.82 | 5.24 | 3.74 | 4.49 |
| | SE (bpm) | 5.60 | 5.75 | 7.31 | 5.83 | 5.79 | **5.48** | 7.40 | 6.75 | 5.54 | 6.18 |
| | r | 0.78 | 0.81 | 0.44 | 0.74 | 0.77 | **0.77** | 0.60 | 0.68 | 0.80 | 0.74 |
| Previous method (SNR weighting) | MAE (bpm) | 3.86 | 4.45 | 7.24 | 4.42 | 4.60 | 4.36 | 5.87 | 5.38 | 4.24 | 4.81 |
| | SE (bpm) | **5.07** | 6.23 | 8.00 | 6.31 | 6.36 | 5.93 | 7.31 | 6.91 | 5.17 | 6.52 |
| | r | **0.84** | 0.76 | 0.30 | 0.69 | 0.69 | 0.71 | 0.61 | 0.66 | 0.74 | 0.70 |
| Proposed Method (Novel Weighting) | MAE (bpm) | **3.74** | **3.83** | **5.65** | **3.57** | **3.87** | 3.99 | **5.25** | **4.89** | **3.44** | **4.17** |
| | SE (bpm) | 5.13 | **5.34** | **6.79** | **5.29** | **5.47** | 5.61 | **6.51** | **6.30** | 5.17 | **5.76** |
| | r | 0.83 | **0.85** | **0.52** | **0.80** | **0.80** | 0.75 | **0.72** | **0.75** | **0.83** | **0.79** |

Table 2. Performance of proposed method as compared to benchmark methods.

*The table shows the performance comparison of the proposed method and the chosen benchmark methods. The metrics shown are Mean Absolute Error (MAE), Standard Deviation of Error (SE) and correlation coefficient (r).*

| Pre-processing | Skin Type | | | Recording Condition | | | | Camera viewpoint | | Overall |
|---|---|---|---|---|---|---|---|---|---|---|
| | Light | Medium | Dark | 3200K | 5600K | Room Lighting | Talking | Front | Lower | |
| Previous method (SNR weighting) | 0.08 | -0.31 | -1.04 | -0.51 | -0.36 | -0.36 | -0.05 | -0.14 | -0.50 | -0.32 |
| Proposed Method (Novel Weighting) | 0.20 | 0.31 | 0.55 | 0.35 | 0.37 | 0.00 | 0.58 | 0.35 | 0.30 | 0.32 |

Table 3. Performance improvement with respect to facial aggregation benchmark of the previous (SNR weighting) method and the proposed method.

*The metric shown is the Mean Absolute Error (MAE) improvement, in bpm.

**FIGURE LEGENDS**

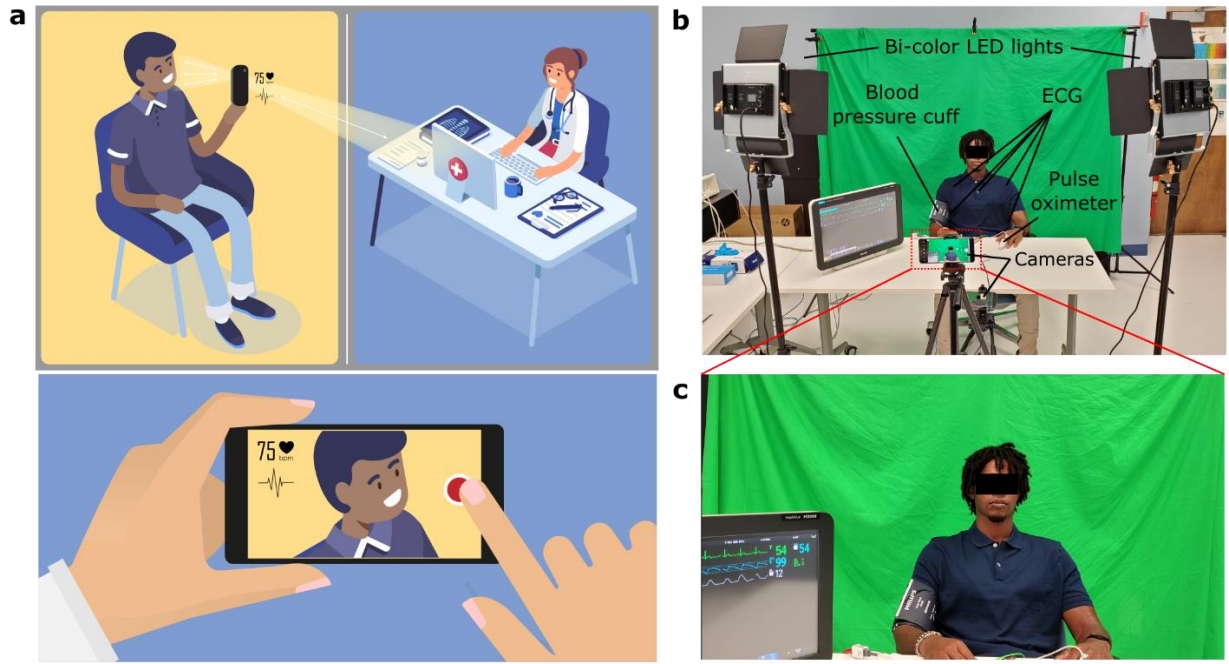

**Figure 1. Constructing a diverse remote vital sign monitoring dataset with a focus on telemedicine applications**. **a.** Cartoon schematic depicting the long-term telemedicine application for the proposed camera-based heart rate estimation. **b.** Experimental setup employed during the construction of the VITAL dataset. Two bi-color LEDs are used for controlled illumination of the subject, and laboratory tube LEDs are used for ambient illumination. The Philips IntelliVue MX800 patient monitor is utilized for ground truth vital sign monitoring. Two smartphone cameras at differing viewing angles capture video of the subject. **c.** Example frame from video captured by the smartphone camera. The subject wears a blood pressure cuff, 5-ECG leads, and a finger pulse oximeter, which is connected to the MX800 unit. Written consent was obtained from the subject for using their image in the publication.

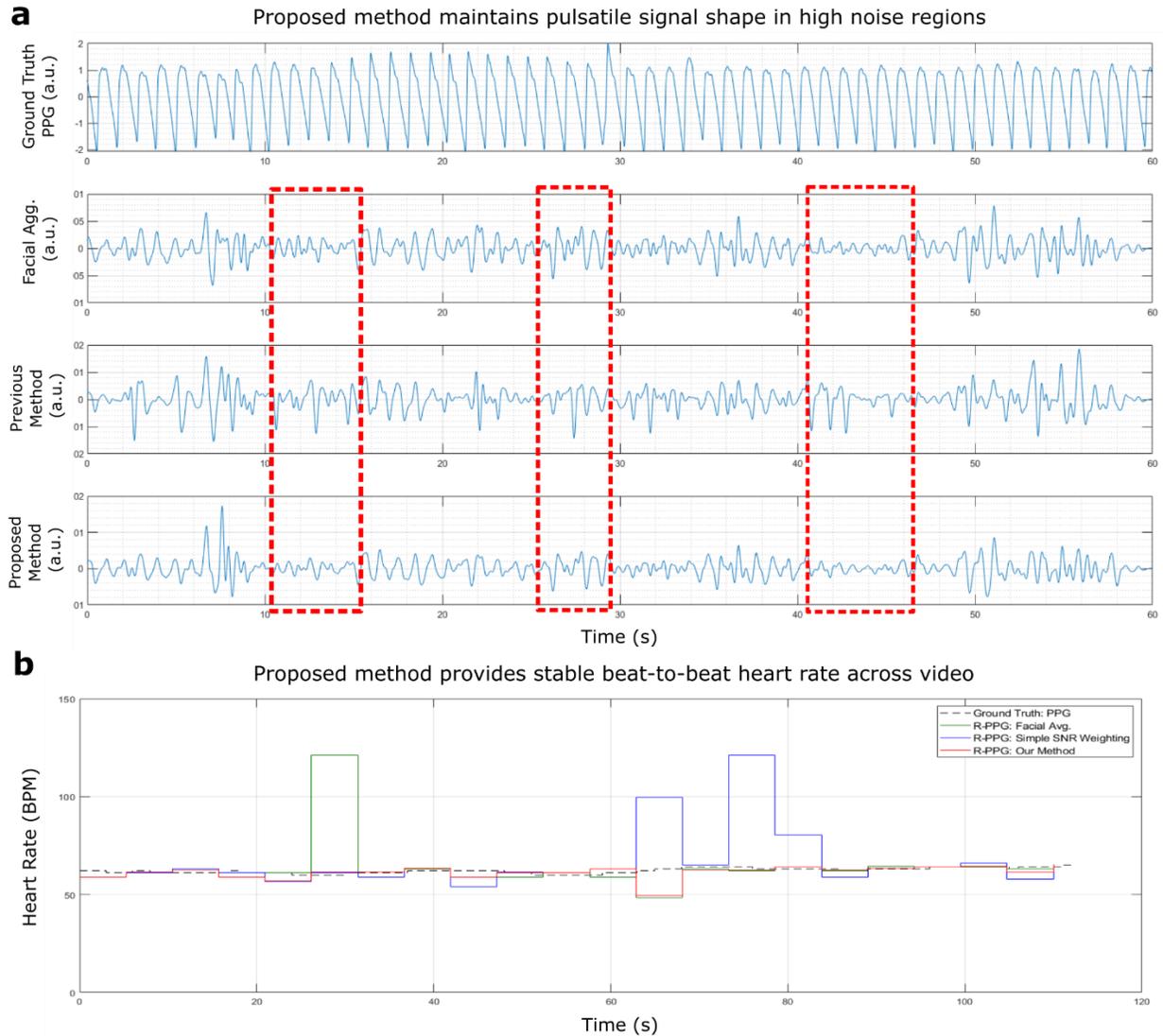

**Figure 2. The proposed method qualitatively recovers the pulsatile signal in a more stable manner compared to prior methods. a.** Example pulsatile waveforms, including the ground truth PPG, facial aggregation r-PPG, previous method's (SNR weighting) r-PPG, and the proposed method's (novel weighting) r-PPG waveform (labelled from top to bottom). The dashed red windows show noisy regions where the r-PPG signal deteriorates. The proposed method maintains pulsatile signal shape, with pulsatile peaks seen more clearly and distinctly. **b.** Beat-to-beat heart rate numerics over time are captured by the proposed method in a more stable manner, consistently staying within 5 bpm of the ground truth PPG.

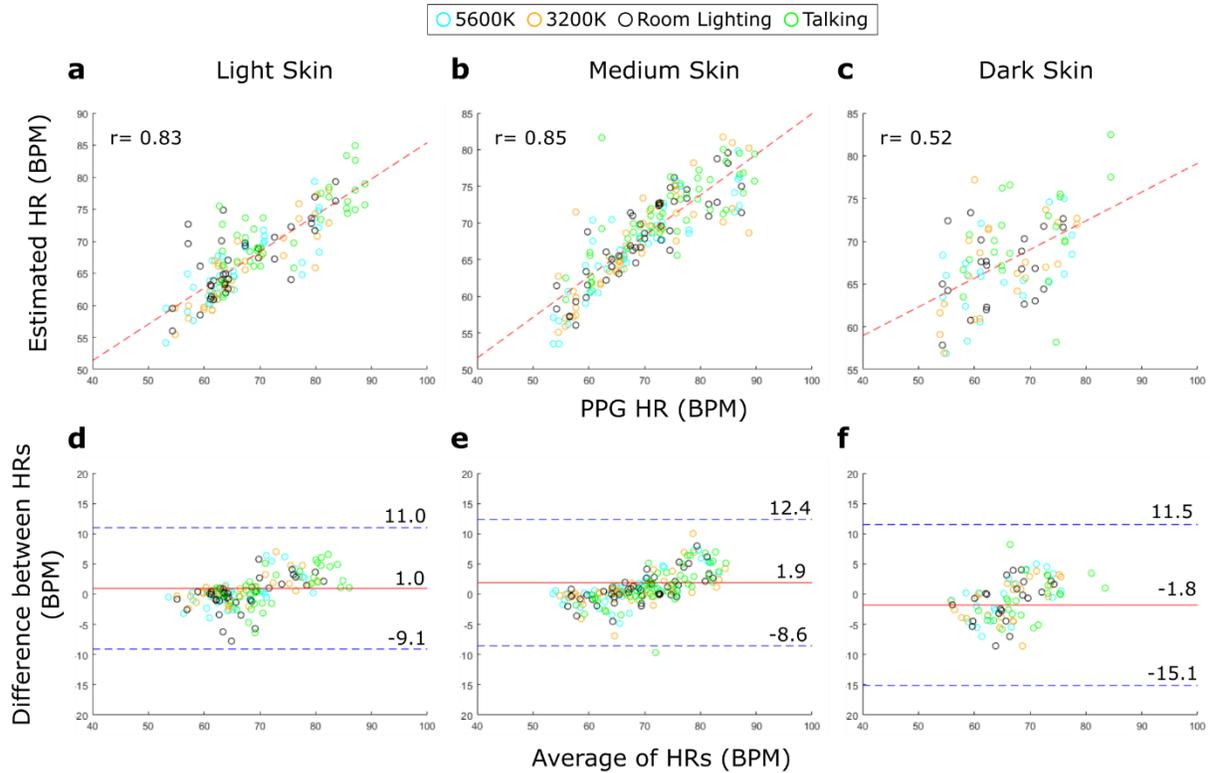

**Figure 3**. **Scatter and Bland Altman plots for proposed heart rate recovery method, varied across skin tone categories.** The label shows a marker for each video recording condition. Performance gap between skin tones is still present, however, is reduced in comparison to previous works (see Table 2). **a-c.** Scatter plots for different skin types. The proposed method shows moderate to strong correlation with respect to ground truth heart rates from the MX800, denoted by the Pearson Correlation Coefficient *r*, across all skin types. **d-f.** Bland-Altman plots for different skin types. The bias (m) is shown by the middle solid red line, and the limits of agreement (LoA = m ± 1.96 σ) by the upper and lower dotted blue lines.

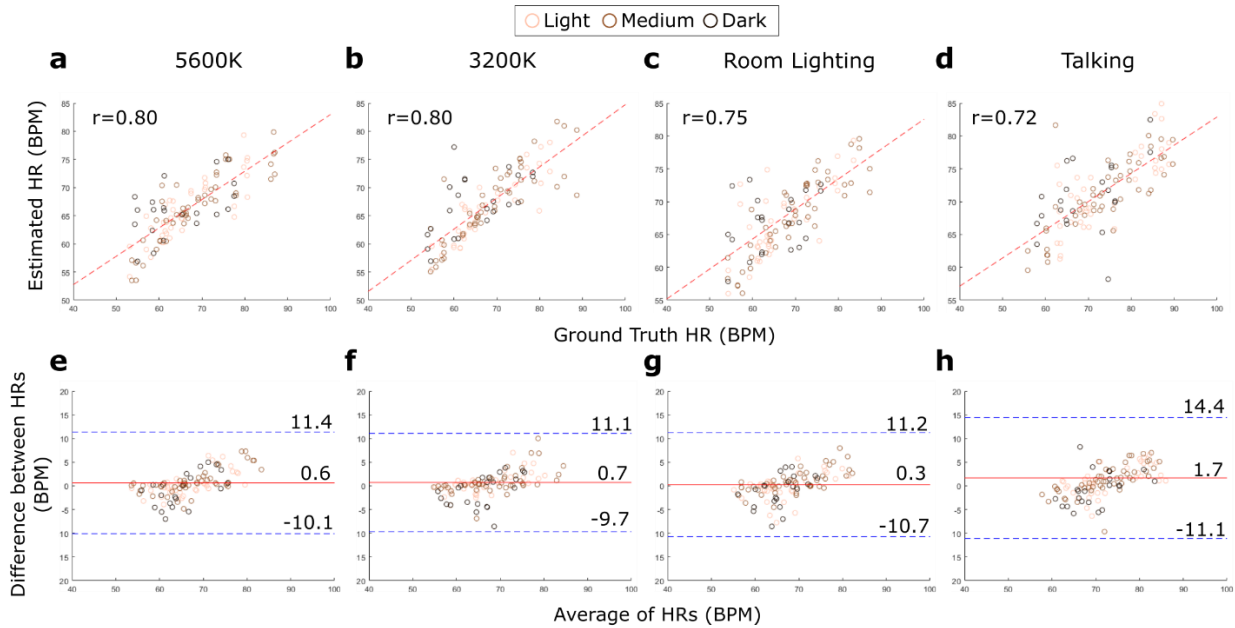

**Figure 4. Scatter and Bland Altman plots for proposed heart rate recovery method, varied across scene condition categories.** The label shows a marker for each skin type. **a-d.** Scatter plots for different recording conditions. The proposed method shows strong correlation with respect to ground truth heart rates from the Philips IntelliVue MX800, denoted by the Pearson Correlation Coefficient *r*, across all recording conditions. **e-h.** Bland-Altman plots for different recording conditions. The bias (m) is shown by the middle solid red line, and the limits of agreement (LoA = m ± 1.96 σ) by the upper and lower dotted blue lines.

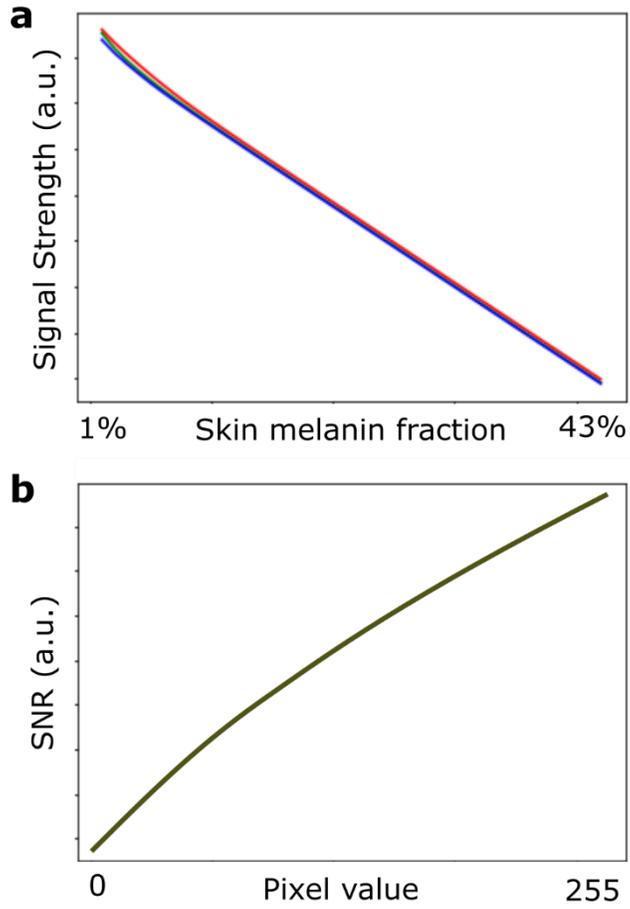

**Figure 5. Theoretical analysis links skin melanin fraction to signal characteristics. a.** Plot of signal strength for biophysical PPG signal for Red, Green and Blue channels, varying with skin melanin fraction. **b.** Plot of signal to noise ratio for a typical camera as a function of increasing pixel intensity.

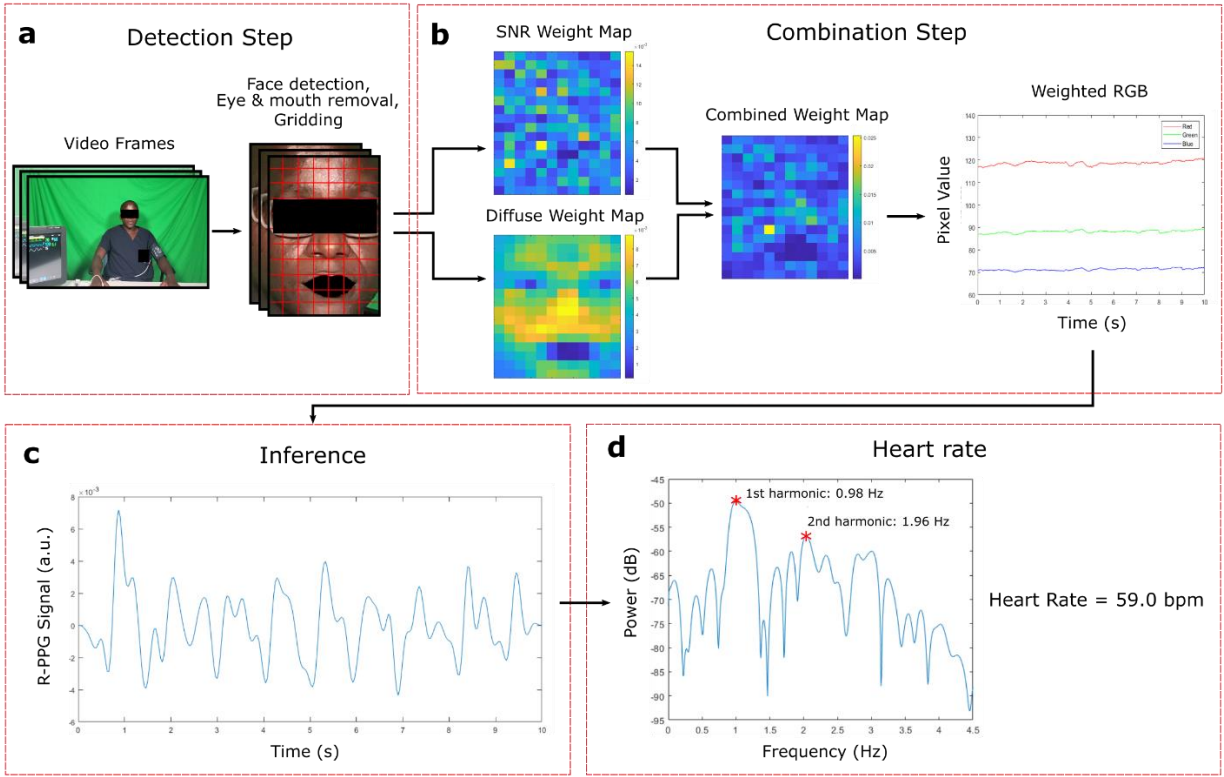

**Figure 6. The proposed heart rate estimation algorithm consists of four steps.** The proposed novelty in the combination step of the pipeline incorporates skin diffuse information weighting, in addition to SNR weighting in RGB space, to achieve robust r-PPG performance across skin tones. Written consent was obtained from the subject for using their image in the publication.

**SUPPLEMENTARY MATERIAL**

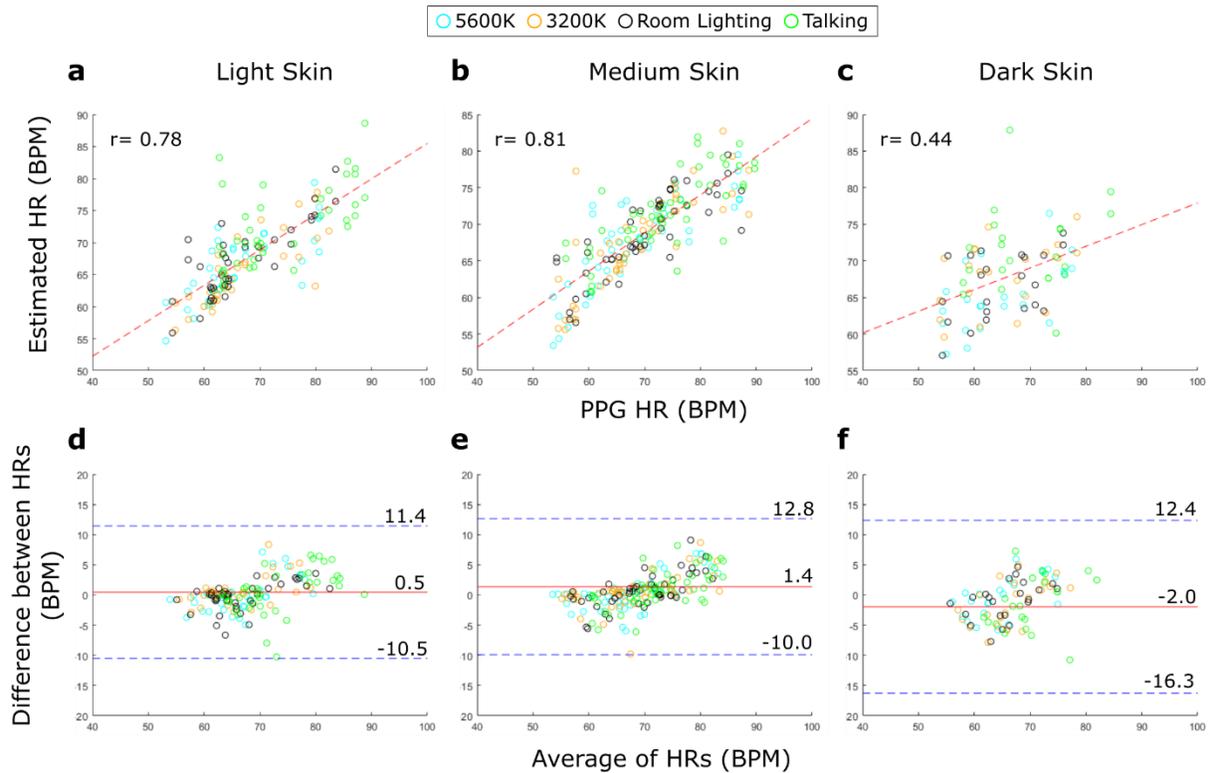

**Supplementary Figure 1**. **Scatter and Bland Altman plots for facial aggregation method, varied across skin tone categories.** The label shows a marker for each video recording condition. **a-c.** Scatter plots for different skin types highlighting the correlation between estimated and ground truth heart rate, denoted by the Pearson Correlation Coefficient *r*. **d-f.** Bland-Altman plots for different skin types. The bias (m) is shown by the middle solid red line, and the limits of agreement (LoA = m ± 1.96 σ) by the upper and lower dotted blue lines.

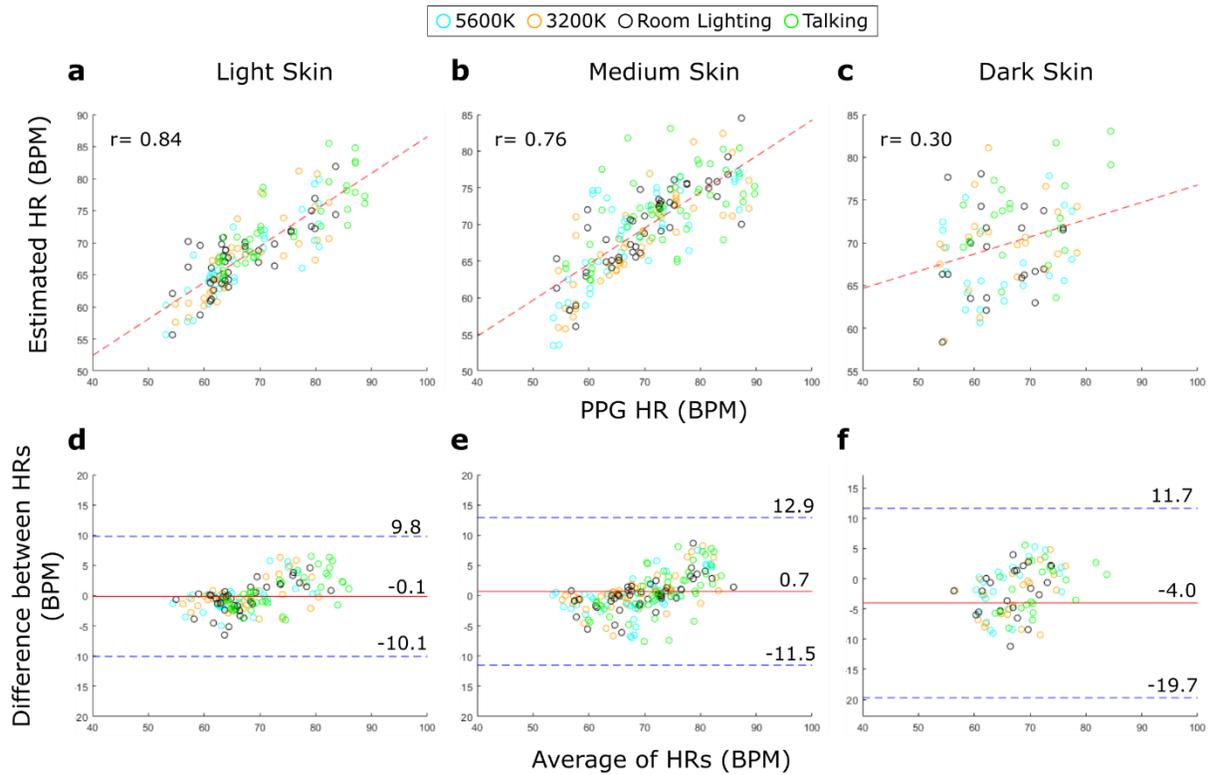

**Supplementary Figure 2**. **Scatter and Bland Altman plots for SNR weighting method, varied across skin tone categories.** The label shows a marker for each video recording condition. **a-c.** Scatter plots for different skin types highlighting the correlation between estimated and ground truth heart rate, denoted by the Pearson Correlation Coefficient *r*. **d-f.** Bland-Altman plots for different skin types. The bias (m) is shown by the middle solid red line, and the limits of agreement (LoA = m ± 1.96 σ) by the upper and lower dotted blue lines.

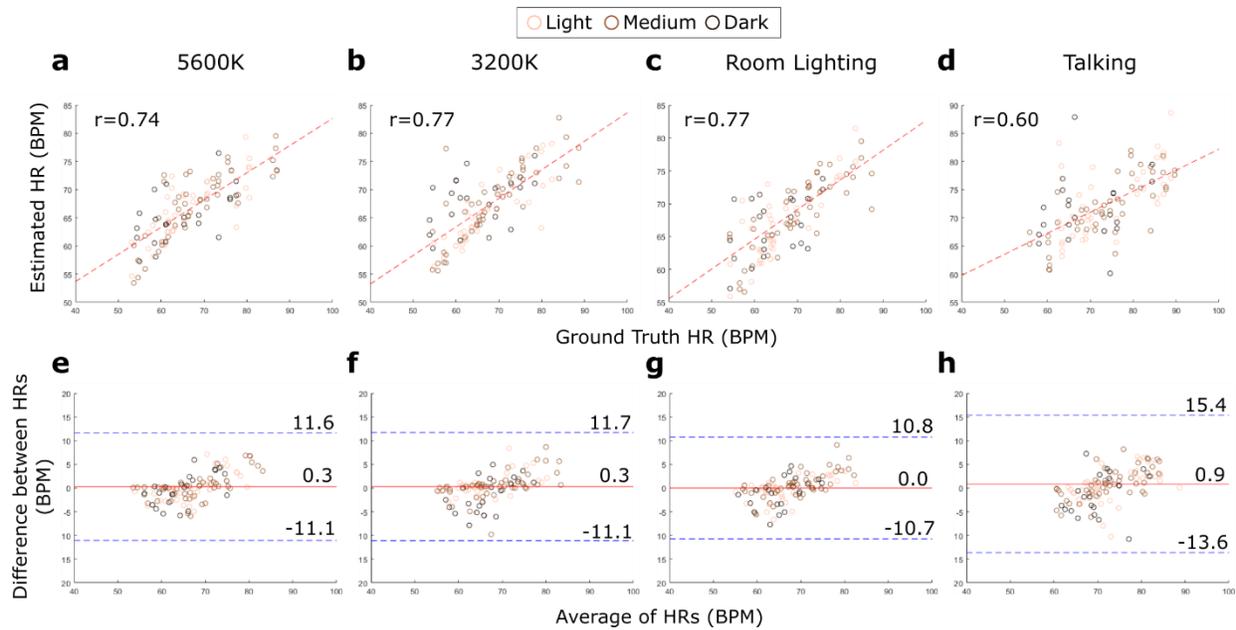

**Supplementary Figure 3. Scatter and Bland Altman plots for facial aggregation method, varied across scene condition categories.** The label shows a marker for each skin type. **a-d.** Scatter plots for different recording conditions highlighting the correlation between estimated and ground truth heart rate, denoted by the Pearson Correlation Coefficient *r*. **e-h.** Bland-Altman plots for different recording conditions. The bias (m) is shown by the middle solid red line, and the limits of agreement (LoA = m ± 1.96 σ) by the upper and lower dotted blue lines.

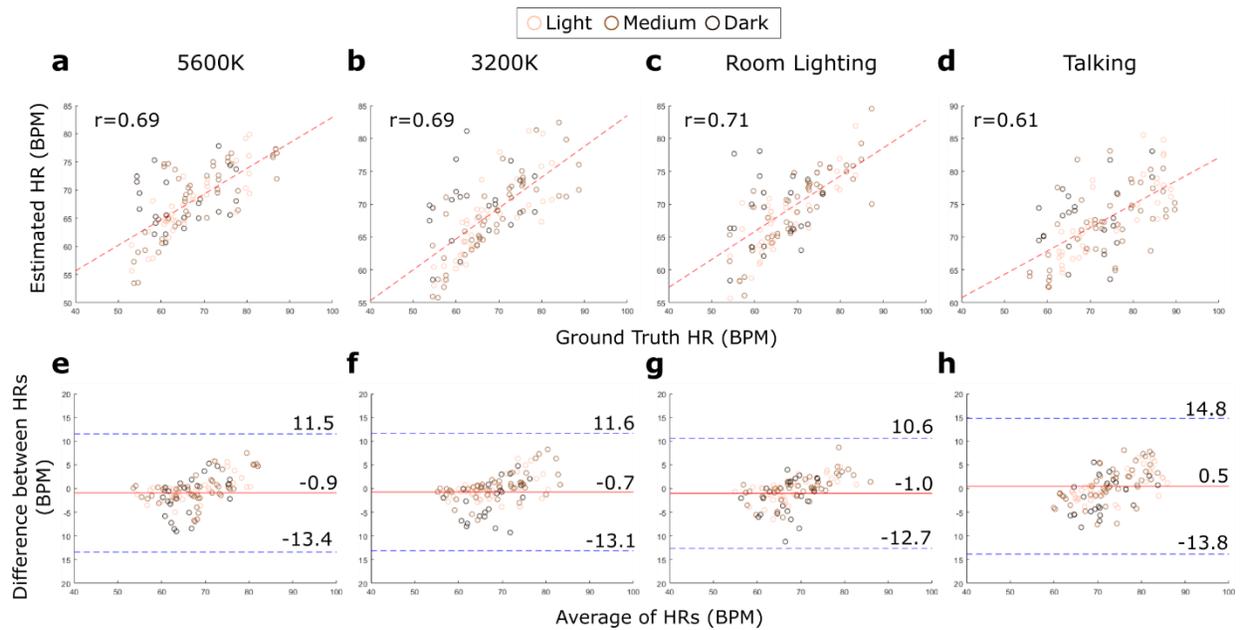

**Supplementary Figure 4. Scatter and Bland Altman plots for SNR weighting method, varied across scene condition categories.** The label shows a marker for each skin type. **a-d.** Scatter plots for different recording conditions highlighting the correlation between estimated and ground truth heart rate, denoted by the Pearson Correlation Coefficient *r*. **e-h.** Bland-Altman plots for different recording conditions. The bias (m) is shown by the middle solid red line, and the limits of agreement (LoA = m ± 1.96 σ) by the upper and lower dotted blue lines.

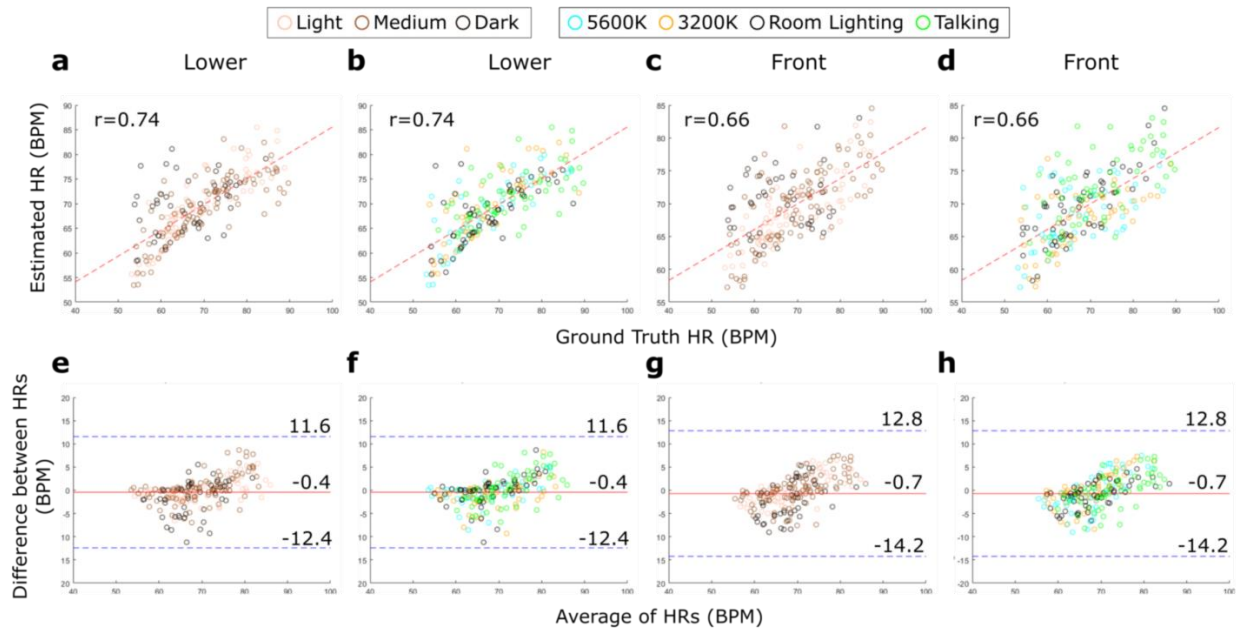

**Supplementary Figure 5**. **Scatter and Bland Altman plots for proposed heart rate recovery method's dependence on camera angle, varied across skin tone categories and recording conditions. a-b.** Scatter plots for the *lower* camera angle, varying across skin tone categories and recording conditions, respectively. **c-d.** Scatter plots for the *front* camera angle, varying across skin tone categories and recording conditions, respectively. **e-f.** Bland Altman plots for the *lower* camera angle, varying across skin tone categories and recording conditions, respectively. **c-d.** Bland Altman plots for the *front* camera angle, varying across skin tone categories and recording conditions, respectively. For all Bland Altman plots, the bias (m) is shown by the middle solid red line, and the limits of agreement (LoA = m ± 1.96 σ) by the upper and lower dotted blue lines.

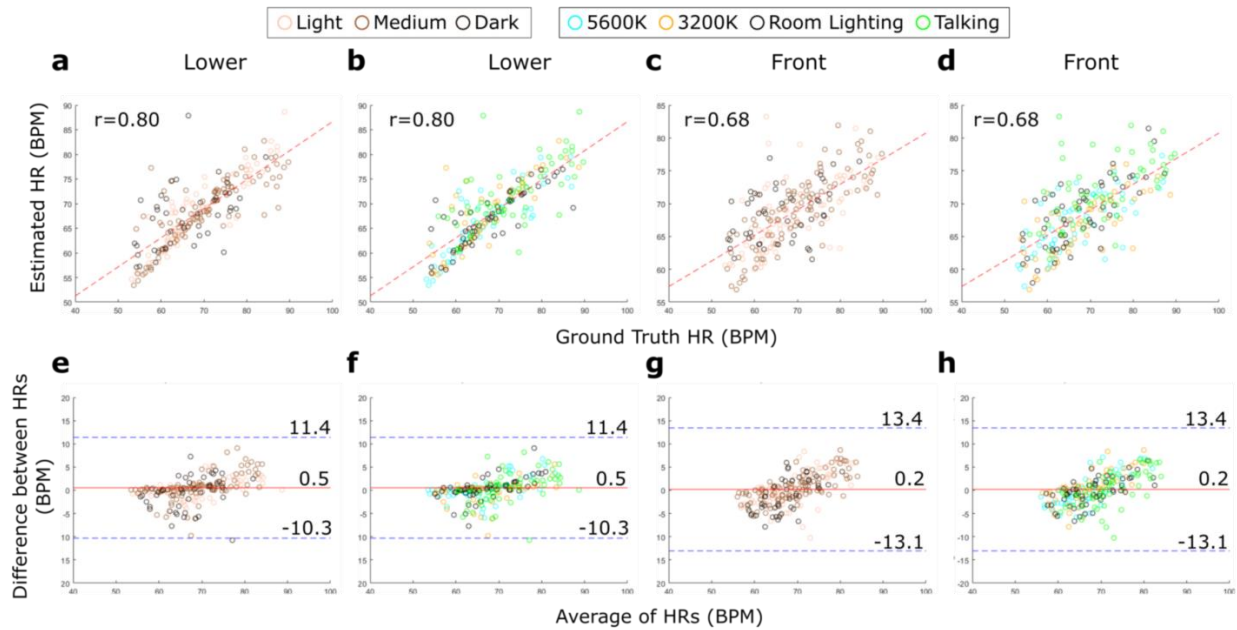

**Supplementary Figure 6**. **Scatter and Bland Altman plots for the facial aggregation method's dependence on camera angle, varied across skin tone categories and recording conditions. a-b.** Scatter plots for the *lower* camera angle, varying across skin tone categories and recording conditions, respectively. **c-d.** Scatter plots for the *front* camera angle, varying across skin tone categories and recording conditions, respectively. **e-f.** Bland Altman plots for the *lower* camera angle, varying across skin tone categories and recording conditions, respectively. **c-d.** Bland Altman plots for the *front* camera angle, varying across skin tone categories and recording conditions, respectively. For all Bland Altman plots, the bias (m) is shown by the middle solid red line, and the limits of agreement (LoA = m ± 1.96 σ) by the upper and lower dotted blue lines.

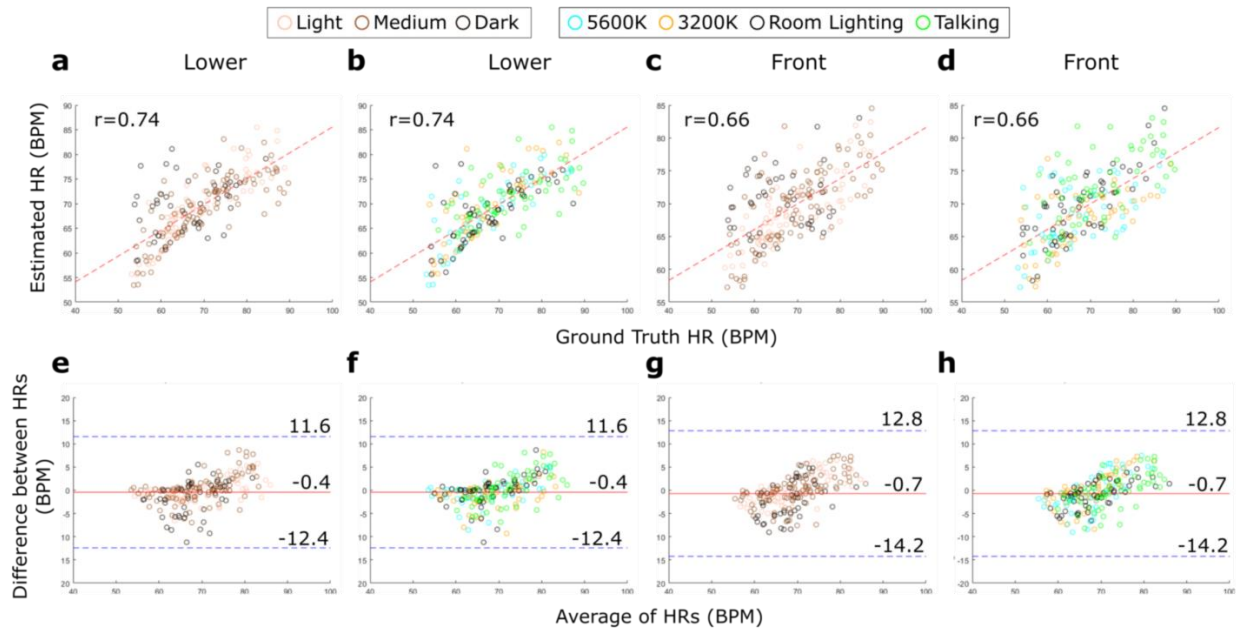

**Supplementary Figure 7**. **Scatter and Bland Altman plots for the SNR weighting method's dependence on camera angle, varied across skin tone categories and recording conditions. a-b.** Scatter plots for the *lower* camera angle, varying across skin tone categories and recording conditions, respectively. **c-d.** Scatter plots for the *front* camera angle, varying across skin tone categories and recording conditions, respectively. **e-f.** Bland Altman plots for the *lower* camera angle, varying across skin tone categories and recording conditions, respectively. **c-d.** Bland Altman plots for the *front* camera angle, varying across skin tone categories and recording conditions, respectively. For all Bland Altman plots, the bias (m) is shown by the middle solid red line, and the limits of agreement (LoA = m ± 1.96 σ) by the upper and lower dotted blue lines.